\documentclass[twocolumn,showpacs,pra,aps]{revtex4}
\usepackage{graphicx}

\begin{document}

\title{Entanglement and Time}

\author{Antoine Suarez}
\address{Center for Quantum Philosophy, P.O. Box 304, CH-8044 Zurich, Switzerland\\
suarez@leman.ch, www.quantumphil.org}

\date{November 1, 2003}

\begin{abstract}

It is argued that recent experiments testing Multisimultaneity
prove that quantum entanglement occurs without the flow of time.
Bohmian mechanics cannot be considered a real temporal
description.\\

\footnotesize\emph{Key words}: Quantum entanglement, nonlocality,
nontemporality, before-before experiments, Einstein's local
realism, Bohm's theory, multisimultaneity.

\end{abstract}

\pacs{03.65.Ta, 03.65.Ud, 03.30.+p, 03.67.Hk}

\maketitle

\section {Introduction}

Quantum Mechanics predicts correlated outcomes in space-like
separated regions for experiments using two-particle entangled
states  \cite{jb64}. Suppose one of the measurements produces the
value $\rho$ ($\rho\in\{+,-\}$), and the other the value $\sigma$
($\sigma\in\{+,-\}$). According to Quantum Mechanics the
probability $Pr(\rho, \sigma)$ of getting the joint outcome
$(\rho,\sigma)$ depends on the choice of the phase parameters
characterizing the paths or channels uniting the source and the
detectors; depending on the value of the phases, the quantity
$Pr(\rho, \rho)$ oscillates between the situation of perfect
correlation ($Pr(\rho, \rho)=1$), and that of perfect
anticorrelation ($Pr(\rho, \rho)=0$). Moreover $Pr(\rho, \sigma)$
is completely independent of any time ordering of the events, i.e.
in the calculation of $Pr(\rho, \sigma)$ the quantum formalism
does not bother about whether the outcome $\sigma$ occurs before
$\rho$, and $\rho$ depends on $\sigma$, or conversely.

In this article we introduce the conditional probability
$Pr(\sigma|\rho)$ as the probability that one of the particles
produces the value $\sigma$ provided the other produces the value
$\rho$, and establish a correspondence between the different
meanings one can attribute to this quantity and the different
views about quantum entanglement. We argue that recent experiments
demonstrate that the dependence expressed by $Pr(\sigma|\rho)$
actually exists but is in principle unobservable and does not
correspond to any real time ordering.

\section {Nonlocal and nontemporal Quantum Mechanics}

To understand the implications of quantum entanglement, it is
convenient to consider first the case of experiments with
entangled photon pairs and \emph{time-like} separated measurements
\cite{ptjr94}. Suppose the outcome $\rho$ occurs in time before
$\sigma$. Even if standard Quantum Mechanics does not bother at
all about which outcome depends on which one, in the case of
\emph{time-like} separated measurements it seems obvious that the
measurement occurring before in time $\rho$, is independent of
that occurring after $\sigma$: Indeed, in this situation, from any
point of view, the after event $\sigma$ does not exist when the
before one $\rho$ takes place. Moreover one could even arrange
that the outcome $\rho$ in side 1 of the setup determines
classically (through light signals) the phase in side 2 and,
thereby, the outcome $\sigma$ in side 2. Experiments with
time-like separated measurements show that the quantum
correlations are compatible with the idea of ordered events, and
even with time-ordered ones. In such experiments the correlations
reveal the dependence of the later outcome on the first one, and
since the events lie time-like separated this dependence can be
considered a time-ordered causal link.

Consider now the case of \emph{space-like} measurements. Here
there is no compulsory ordering, and in principle two different
descriptions are possible:

One can consider that $\rho$ depends on $\sigma$ and describe
things according to the following equation:

\begin{equation}
{Pr(\rho, \sigma)}=Pr(\sigma|\rho) \sum_{\sigma} Pr(\rho,
\sigma)\label{Pr12}
\end{equation}

\noindent where the conditional probability $Pr(\sigma|\rho)$
expresses the dependence of the outcome $\sigma$ on $\rho$.

Or one can alternatively consider  that $\rho$ depends on $\sigma$
and the correlations are worked out according to the following
equation:

\begin{equation}
Pr(\rho, \sigma)= Pr(\rho|\sigma)\sum_{\rho} Pr(\rho, \sigma)
\label{Pr21}
\end{equation}

where the conditional probability $Pr(\rho|\sigma)$ expresses the
dependence of the outcome $\rho$ on $\sigma$.

One can hardly deny that the very fact of the correlations reveals
a dependence actually existing in the world:``Correlations cry out
for explanation'' \cite{jb64}, two events cannot be correlated if
each of them takes place quite at random. Therefore it is
astonishing that Quantum Mechanics only speaks about joint
probabilities, and says nothing about conditional probabilities
expressing the dependence between the outcomes. This silence
protects two deep secrets.

The first one is that of Nonlocality, which John Bell unveiled.
Quantum mechanics implies that the term $Pr(\sigma|\rho)$ (or
$Pr(\rho|\sigma)$) corresponds to a faster than light link between
outcomes. Nonlocality has been tested and did prevail against
Einstein local realism (see section III below).

The second secret is Nontemporality, unveiled by recent research.
The dependence expressed by the term $Pr(\sigma|\rho)$ (or
$Pr(\rho|\sigma)$), actual though it is, it doesn't correspond to
any real temporal ordering, it doesn't have any observable
counterpart. The correlations arise without the flow of time.
Nontemporality has been tested and did prevail against
Multisimultaneity (see section V below).

\section {Testing nonlocality vs Einstein's local realism}

According to the local view of Relativity one has to exclude any
link faster than light and, therefore, any direct connection
between space-like separated regions. This means that the
dependence expressed by $Pr(\sigma|\rho)$ or $Pr(\rho|\sigma)$
doess not arise from a direct link between the two measurements,
but originate from some common cause in the absolute past of both
measurements. Apparently this was the way Einstein thought, and
concluded that the quantum mechanical description of the physical
reality cannot be considered complete \cite{epr}.

However John Bell showed that if one only admits relativistic
local causality (causal links with $v\leq c$), the correlations
occurring in two-particle experiments should fulfill clear
locality conditions (``Bell's inequalities''), which by contrast
are violated by Quantum Mechanics (Bell's theorem) \cite{jb64}.
Bell experiments conducted in the past two decades, in spite of
their loopholes, suggest a violation of local causality:
statistical correlations are found in space-like separated
detections; violation of Bell's inequalities ensure that these
correlations are not pre-determined by some common cause in the
past \cite{exp}. Nature seems to behave nonlocally, and Quantum
Mechanics predicts well the observed distributions.

Even if the quantum nonlocality violates Einstein's view that
nothing in nature goes faster than light, it does not lead to any
conflict with Relativity on the level of the observations.
Effectively, the quantum mechanical formalism ensures that the
unobservable ``Bell connections'' cannot be used to phone
faster-than-light \cite{eb78}.

\section {Bohm's theory}

If the correlations are not pre-determined, then there must be
some direct dependence or connection between two space-like
separated events. But what kind of causality does such a direct
dependence involve?

Bohm's theory \cite{dbbh} was the first attempt to cast
nonlocality into a time-ordered causal scheme. It gives up the
relativity of time and uses a unique preferred frame or absolute
time, in which one event is caused by some earlier event by means
of instantaneous action at a distance. The value occurring later
in this preferred frame, say $\sigma$ depends on the value $\rho$
occurring before. In this sense Bohm's action at a distance refers
to a causal link though not a Lorentz-invariant one. The
description makes always the same predictions as Quantum
Mechanics. At the observable level no conflict with relativity
arises.

However, if one tries to cast nonlocal causality into only one
preferred frame it is not more reasonable to connect a ``cause''
event to an ``effect'' event in that frame rather than in some
other frame. Effectively a single preferred frame (``quantum
ether'') is ``experimentally indistinguishable" \cite{jb64}: The
predictions would remain the same if one assumes that the
preferred frame is a virtual entity changing from experiment to
experiment.

One is tempted to think that Bohm introduces absolute time just
because he wishes to justify a causal description, but in the end,
an untraceable ``quantum ether'' is essentially the same as
deciding arbitrarily which event depends on which one \cite{pe99}.
What is more, in the particular case, possible in principle, of
both measurements taking place at exactly the same time in the
preferred frame, the only way of establishing which event depends
on which is by arbitrary decision. Actually, Bohmian Mechanics can
be considered a causal description but not a real temporal one,
and to date it has not lead to experiments capable of
distinguishing it from Quantum Mechanics. Regarding the meaning of
the dependence $Pr(\sigma|\rho)$, in the end Bohmian Mechanics
shares the same view as Quantum Mechanics: The dependence actually
exists but it is essentially unobservable; the ordering expressed
in $Pr(\sigma|\rho)$ does not correspond to any real time
ordering.

\section {Testing Nontemporality vs Multisimultaneity}

As long as one believed (according to Einstein) that there are no
space-like influences, the fundamental temporal notion could not
be other than proper time along a time-like trajectory. But since
Bell experiments did reveal a world consisting in nonlocal
connected events, the ``reasonable'' position in the very spirit
of relativity is to assume time-ordered causality, and describe
the nonlocal links using lines of simultaneity to distinguish
between ``before'' and ``after''.

Indeed, such a description is very well possible in conventional
Bell experiments, in which all apparatuses are standing still in a
laboratory frame. Since the emission time of the photons is not
exactly the same, and the fibers guiding the photons from the
source to the measuring devices do not have exactly the same
length, according to the clock defined by the laboratory's
inertial frame, one of the measurements always takes place before
the other, and the particle arriving later can be considered to
take account of the outcome of the one arriving before. In fact,
this is the way Bell tried to explain things, and, in doing so
(i.e. assuming that $Pr(\sigma|\rho)$ reflects the the time
ordering in the laboratory frame) he came to discover quantum
nonlocality \cite{jb64}. Orderings with one measurement before and
the other after in time are referred to as \emph{before-after} or
\emph{after-before} timings. In experiments with all measuring
devices at rest, it is possible to explain quantum correlations
through time-ordered (nonlocal) causality.

But what about experiments with moving apparatuses in which
several relevant frames are involved? In this case, different
clocks watch the arrival times, and what is ``after'' according to
the laboratory clock may become ``before'' according to one moving
clock. Then, it is possible to define other time orderings: If
each measuring device in its own reference frame is the first to
select the output of photons, we have \emph{before-before} timing.
If each measuring device in its own reference frame selects the
photon output after the other, we have \emph{after-after} timing.
Is it also possible to give a time-ordered causal explanation for
relativistic experiments using apparatuses in motion?

Recently, an alternative nonlocal description called
Multisimultaneity has been proposed, which extends the
time-ordered description to experiments with \emph{before-before}
and \emph{after-after} timings \cite{asvs97,as97,as00.1}. This
description imbeds nonlocality in a real relativistic time
ordering by using several relevant frames. The main motivation of
such a proposal is to create an experiment allowing us to decide
whether nonlocal influences can be measured by means of several
real clocks.

Multisimultaneity assumes basically Bohm's idea of time-ordered
nonlocal causality but uses real and well defined inertial frames
to describe the causal links: the inertial frames of the measuring
devices. The basic assumption of Multisimultaneity is that the
decision about the output port by which a photon leaves a
choice-device takes account of all the local and nonlocal
information available within the inertial frame of this
choice-device, at the instant the particle strikes it. The
dependence expressed by $Pr(\sigma|\rho)$ is tied to an
experimentally well defined inertial frame and corresponds to a
real time ordering. Within each choice-device's frame the causal
links always follow a well defined chronology, one event never
depending on some future event.

As said, in the conventional Bell experiments there is only one
relevant frame, and in it one of the choices, say, $\rho$ takes
place always before the other $\sigma$, and the particle arriving
later takes account of the decision of that arriving first, just
as indicated in equation (\ref{Pr12}). Therefore,
Multisimultaneity bears the same predictions as Quantum Mechanics.
In experiments with all devices at rest it is possible to give a
causal explanation in which the ordering of the events fits with
the time ordering in the laboratory frame.

Consider now experiments in which the choice-devices are in motion
in such a way that each of them, in its own reference frame, is
first to select the output of the photons (\emph{before-before}
timing). Then, each particle's choice will become independent of
the other's, and according to Multisimultaneity the nonlocal
correlations should disappear. By contrast Quantum Mechanics
requires that the particles stay nonlocal correlated independently
of any timing, even in such a \emph{before-before} situation
\cite{as00.1}. This means that \emph{before-before} experiments
are capable of acting as standard of time-ordered nonlocality
(much as Bell's experiments act as standard of locality): if
timing-independent Quantum Mechanics prevails, nonlocality cannot
be imbedded in a relativistic chronology; if Quantum Mechanics
fails, there is a time ordering behind the nonlocal correlations,
and proper time along a time-like trajectory is not the only
temporal notion \cite{ip98}.

As a matter of fact, the conventional Bell experiments did not
test in any way whether or not entanglement depends on the timing
of the measurements in relativistic setups. Therefore, also taking
nonlocality for granted, it was necessary to make experiments with
moving measuring devices.

This means that if one invokes a description assuming the quantum
collapse one has to put the detectors in motion, and if one
invokes a description without collapse (in the line of Bohm's
theory) the monitored beam splitters. We stress that in case of
the beam splitters this frame is unambiguously defined by the
velocity corresponding to the Doppler-shift of the reflected
photons \cite{as00.1}.

Multisimultaneity proposes feasible experiments using moving
choice-devices \cite {asvs97,as97,as00.1}, which are of interest
in the general context of physical situations involving several
observers in relative motion \cite{pe01}, specially to the aim of
testing the timing-independence of the quantum probabilities. All
four possible relativistic timings have been experimentally
tested. In February 2000 experiments using moving detectors showed
results in agreement with Quantum Mechanics \cite{zbgt}, i.e. no
disappearance of the correlations was observed with
\emph{before-before} and \emph{after-after} timing. More recently,
acousto-optic modulators have made it possible to perform
experiments with moving beam-splitters \cite{as00.1}. The results
obtained in June 2001 uphold the predictions of Quantum Mechanics
too \cite{szsg}.

\section {Other temporal schemes}

We discuss now two other ways of imbedding entanglement in a
temporal scheme, which have strong implications for the
understanding of causality and involve peculiar experimental
conditions.

Ph. Eberhard's theory uses influences propagating at finite
velocity $V>c$. Such a theory has been proposed by \cite{eb89}.
The value of the new constant $V$ is not given, but possible
experiments are described, which would allow us to establish it
providing they prove standard Quantum Mechanics wrong. Since the
preferred frame is in principle experimentally distinguishable, it
would define a real universal clock and, therefore, the assumed
causality is a temporal one. However, the experiments proposed
would not be capable of discarding the preferred-frame
description: upholding of the quantum mechanical predictions would
simply establish a lower bound for the speed $V$ of the
superluminal influences causing the correlations. As Eberhard
himself shows, the influences propagating at a finite speed $V>c$
could be used for superluminal communication \cite{eb89}. Thus,
the belief in a \emph{real} preferred frame, if it is serious,
should lead to \emph{real} experiments aiming to demonstrate
superluminal signaling \cite{ophoc}. But we fear the fact that the
theory cannot be falsified when tested against Quantum Mechanics
is preventing physicists from performing the proposed experiments.

O. Costa de Beauregard sticks to Einstein's postulate that any
causal link has a Lorentz invariant counterpart in the formalism,
and proposes to explain entanglement by means of influences
propagating along the light-cone but backwards in time
(retrocausation):  The measurement of one of the particles does
not influences directly the other particle but the source through
backwards causation \cite{co}.

In reality "backwards causation" is the strongest form of the bias
that causality is always bond to the flow of time (one could also
say the bias that all causal links have to be essentially Lorentz
invariant): If one cannot bind causality to the time flowing
forwards, one binds it to the time flowing backwards! This view
seems to bear a big problem in situations in which the two
measurements lie time-like separated, for instance the measurement
of photon 2 lies in the future light cone of the measurement of
photon 1. Indeed, in this case "backwards causation" means that an
event occurring at time T (the measurement of photon 1) is
determined by another event (the measurement of photon 2), which
at time T has no existence at all in any possible inertial frame
(of the source or other).

As far as such a view does not imply observable effects there is
no problem in maintaining it: it is merely another way to say the
same as Quantum Mechanics, as it is the case also for Bohm's
theory. Nevertheless Costa de Beauregard pretends that backwards
causation should have observable implications in experiments
involving psi-subjects enjoying capacities as telepathy. In
experiments with entangled particles, such psi-subjects are
supposed to be capable of influencing on purpose the probability
to get a particular outcome, say +, on one side of the setup, and
because entanglement they would also change faster than light the
corresponding probability on the other side. This would imply the
possibility of signaling faster than light and, therefore, of
changing the past. As it seems, there is work in progress to test
Costa de Beauregard's prediction. Additionally to the difficulty
of establishing which subjects can be considered to posses
psi-capacities, the question is whether the corresponding
experiments will be acknowledged as valuable tests beyond the
psi-community.

\section {Conclusion}

The results of experiments with moving measuring devices mean that
the quantum correlations are caused regardless of any relativistic
chronology: entangled photons run afoul of the relativity of time,
Einstein's frames have no effect on ``spooky action'', even though
we cannot use this fact to establish an absolute time.

So what Quantum Mechanics actually implies is that in case of
space-like separated measurements the dependence the correlations
reveal does not correspond to any real time ordering and,
consequently, is not tied to any experimentally distinguishable
frame. In spite of the different orderings the equations
(\ref{Pr12}) and (\ref{Pr21}) bear the same joint probability of
getting the outcome $(\rho,\sigma)$, i.e., the measurable quantity
$Pr(\rho, \sigma)$ predicted by Quantum Mechanics, which is
independent of any ordering and timing. To produce the
correlations, nature can choose between the ordering assumed in
(\ref{Pr12}) and that in (\ref{Pr21}) but its choice has no
observable consequence at all, and it is not possible, even in
principle, to distinguish which measurement is the independent and
which the dependent one. Any observer who would record the
temporal sequence of the outcomes in his own frame, and assume
that the outcome later in time depends on the earlier one
according to one of the rules (\ref{Pr12}) or (\ref{Pr21}), would
make the same predictions as Quantum Mechanics.

The term $Pr(\rho, \sigma)$ expressing the measurable joint
probability of getting a given outcome is perfectly
Lorentz-invariant. As for the quantities $Pr(\sigma|\rho)$ and
$Pr(\rho|\sigma)$ expressing the dependence between the events
they are certainly not Lorentz invariant but don't have any
measurable counterpart. Quantum entanglement implies causal links
which are not Lorentz-invariant. Nevertheless such causal links do
not imply any \emph{observable} violation of Lorentz invariance at
all.

The influences allowing us to phone between two separated regions
follow time-like trajectories, and can consistently be described
in terms of ``before" and ``after" by means of real clocks;
Einstein's world contains only such local causal links. The
entanglement bringing about nonlocal correlations is insensible to
space and time, and cannot be described in terms of ``before" and
``after" by means of any set of real clocks. The notion of time
makes sense only in Einstein's world, i.e. along time-like
trajectories. Suppose a physicist could act non-locally and would
like to bring about Bell-correlations; she or he would first
choose one event assigning randomly a value (either $+$ or $-$) to
it, and subsequently would assign a value depending upon the
first, to the second event. Suppose these operations occur without
the flow of time; as Quantum Mechanics seems to mean, and the
experimental results confirm, this is the way things happen in
nature.

In conclusion the experiments testing quantum entanglement rule
out the belief that physical causality necessarily relies on
observable signals. Quantum entanglement supports the idea that
the world is deeper than the visible, and reveals a domain of
existence, which cannot be described with the notions of space and
time. In the nonlocal quantum realm there is dependence without
time, things are going on but the time doesn't seem to pass here.

\section* {Acknowledgements}

I would like to thank Nicolas Gisin, Valerio Scarani, Andr\'e
Stefanov, and Hugo Zbinden for very stimulating discussions, and
the Odier Foundation of Psycho-physics and the L\'eman Foundation
for support.

\end{document}